\documentstyle[prb,aps,epsf,floats,twocolumn]{revtex}

\begin{document}
\draft

\author{Akihisa Koga and Norio Kawakami}
\address{Department of Applied Physics, 
Osaka University, Suita, Osaka 565-0871, Japan}
\title{Mixed-spin cluster expansion for a quasi-one-dimensional 
Haldane system}
\date{\today}
\sloppy
\maketitle

\begin{abstract}
We present a novel mixed-spin cluster expansion method for a 
quasi-one-dimensional Haldane system with bond alternation.
By mapping the $s=1$ antiferromagnetic
spin model on square and cubic lattices to the equivalent 
mixed-spin model, we  study the competition among  
the Haldane, the dimer  and the magnetically ordered phases. 
The  mixed-spin cluster expansion proposed here
 allows us to directly deal with the Haldane 
phase, which may not be reached by standard 
series expansion methods. The phase diagram is determined rather
precisely by making use of an additional symmetry property in 
the effective mixed-spin model introduced.
\end{abstract}

\narrowtext
Low-dimensional spin systems with the spin gap for the excitation spectrum
have been extensively studied since the Haldane conjecture,\cite{Haldane}
which clarified that the gap formation in the 
integer-spin Heisenberg chain  reflects  the topological nature of spins.
Recent extensive experimental and theoretical investigations 
on the stability of the Haldane system against various perturbations
have been providing a variety of interesting topics.
The instability of the spin-gap phase in the 
 $s=1$ spin models has been studied in detail so far for one-dimensional
(1D) systems.
For instance, the effect of the bond alternation is understood 
qualitatively well
by the non-linear sigma model\cite{NLSM},  as well as 
the valence bond solid (VBS) approach.\cite{VBS}
The accurate critical point between the 
 dimer and the Haldane phases has been further obtained by 
the series expansion,\cite{Gelone}
the exact diagonalization,\cite{KatoTanaka,Totsuka}
the quantum Monte Carlo simulations\cite{Yamamoto}
and the density matrix renormalization group (DMRG).\cite{KatoTanaka}
On the other hand, the $s=1$ spin systems 
with the 2D or 3D structures have not been studied so well,
although the effects of the antiferromagnetic 
correlations due to the interchain couplings should be important
for real materials. 
So far, Sakai and Takahashi\cite{Sakai2D} 
investigated a quasi-1D $s=1$ spin system
by combining the mean field  theory with 
the exact diagonalization results for the spin chain,  
and gave a rough estimate for the phase-transition point
to the antiferromagnetic phase.

In  this paper,  we  systematically study how the  Haldane 
and the dimer phases for the $s=1$  antiferromagnetic chain
 are driven to the magnetically ordered phase in 2D and 3D systems
by exploiting  the series expansion techniques.
In particular, we propose  a {\it mixed-spin cluster expansion}  
by mapping the $s=1$ spin model to the equivalent 
mixed-spin model, which  allows us to deal with the Haldane 
phase. This new approach is a realization of the notion of the VBS
 in a perturbation theory.
We determine the phase diagram rather precisely both
for the  2D and 3D cases by 
computing  the spin excitation gap and the staggered susceptibility.

Let us first consider the $s=1$ antiferromagnetic quantum spin system
on a 2D square lattice, 
which is described by the Hamiltonian
\begin{eqnarray}
H&=&\sum_{i,j} \left[ \Gamma_{i} {\bf S}_{i,j}\cdot{\bf S}_{i+1,j}
+J {\bf S}_{i,j}\cdot{\bf S}_{i,j+1}\right],\label{eq:model}
\end{eqnarray}
where $J$ is the interchain coupling
and ${\bf S}_{i,j}$ is the $s=1$ operator at the $(i, j)$-th site 
in the $(x-y)$ plane.
Here we have introduced  the bond-alternation parameter 
$\alpha (0\leq\alpha\leq 1)$  along the $x$ direction, 
$\Gamma_i=1 (\alpha)$ for even (odd) $i$.
All the exchange couplings are assumed to be antiferromagnetic.

We employ the series expansion method developed 
by Singh, Gelfand and Huse.\cite{dimer1}
Since this method combines  
the conventional perturbation theory with the cluster expansion,
it has an advantage to deal with the spin system in higher dimensions 
even for the cases for which  the reliable results are difficult to be obtained
by the exact diagonalization, the DMRG, etc. 
In fact, the series expansion method 
has been successfully applied  to the
2D spin systems with various structures,\cite{dimer1,2Dsystem} 
Kondo lattice,\cite{Kondo} bilayer systems,\cite{bilayer} etc.
However, to apply the series expansion technique
to the present system including the Haldane phase, 
a nontrivial generalization is needed, since a naive cluster expansion 
may not describe the Haldane state. For instance, 
the dimer state is adiabatically
connected to the isolated $s=1$ dimers, but
the Haldane state does not have its analogue in 
the isolated local singlets composed of several $s=1$ spins.
\begin{figure}[htb]
\epsfxsize=7cm
\centerline{\epsfbox{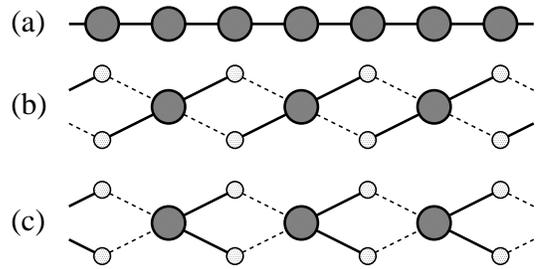}} 
\vspace{0.1cm}
\caption{(a) The $s=1$ spin  chain, which is  decomposed into 
(b)  the mixed-spin chain  in the Haldane phase and  into 
(c) that in the dimer phase. 
Large and small solid circles represent the $s=1$ spin and $s=1/2$ spins, 
respectively. 
The solid lines in (b) and (c) indicate the strong bonds 
which make local singlets whereas 
the dashed lines the weak bonds which are treated perturbatively.
}
\label{fig:chn}
\end{figure}
To overcome this problem, we wish to recall the notion of the VBS\cite{VBS},
which captures the essence of the Haldane-gap formation. To realize
this idea in the series expansion, we first  
divide half of the $s=1$ spins into two $s=1/2$ spins
as schematically shown in Fig. \ref{fig:chn},\cite{comment} and map 
the system to the mixed-spin system which is equivalent to the original model
except for  a trivial isolated excited  mode.
As a starting configuration in the perturbative expansion, 
we can then consider two types of the mixed-spin cluster singlets 
formed by the solid lines in Figs. \ref{fig:chn} (b) and (c).
It is seen that by starting from the  configuration (b)
we can directly deal with the Haldane phase  
since it  has the structure of the Haldane state in the VBS picture,
whereas if  the configuration (c) is chosen,
we  naturally end up with  the standard dimer expansion.
The above mapping thus  gives us an important message 
that {\it  the Haldane phase is
adiabatically connected  to the isolated mixed-spin singlet states in 
Fig. \ref{fig:chn}(b),  
and thereby can be treated by the mixed-spin cluster expansion method.}
The resulting cluster expansion around the isolated  mixed-spin 
singlets  should provide a quite powerful method, 
which enables us to deal with the competition among
 the Haldane phase, the dimer phase and 
 the magnetically ordered phase in 2D and 3D systems.

\begin{figure}[htb]
\epsfxsize=8cm
\centerline{\epsfbox{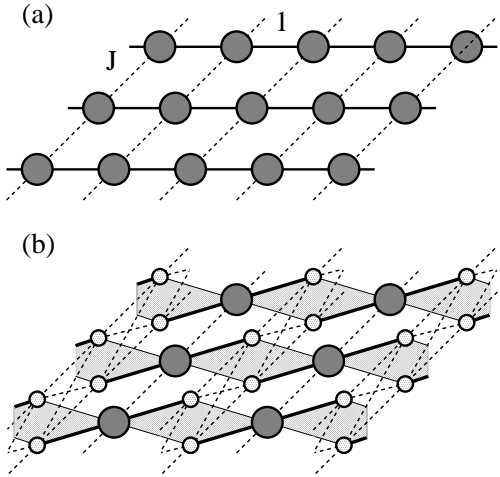}} 
\vspace{0.1cm}
\caption{(a) The 2D $s=1$ spin model and
(b) the corresponding  mixed-spin system.}
\label{fig:2d}
\end{figure}

Let us begin  with the quantum phase transition
between the Haldane phase  and the antiferromagnetic phase in 2D $s=1$ 
spin system with bond alternation.
To this end, we consider the effective 2D mixed-spin system 
shown in Fig. \ref{fig:2d}.
In this figure, the large (small) circle represents the $s=1 (s=1/2)$ spin.
The bold solid, the thin solid and the dashed lines indicate 
the coupling constant $1$, $\lambda$ and $J\lambda$, respectively.
In this figure, the model without bond alternation is drawn for simplicity.
We note that the mixed-spin system reproduces
the original 2D spin system at $\lambda=1$.
To perform the cluster expansion,  
the Hamiltonian is divided into two parts as
$H=\sum {\bf S}_i\cdot{\bf S}_j+\lambda\sum {\bf S}_i\cdot{\bf S}_j$.
The first term is the unperturbed Hamiltonian which stabilizes the 
isolated mixed-spin cluster singlets.
The corresponding mixed-spin cluster has the configuration,
 $1/2\circ 1\circ 1/2$, which is formed by the antiferromagnetic 
couplings $1$ and $\alpha$. 
These isolated clusters have the singlet ground state with the spin gap 
$\Delta=(3\alpha+3-\sqrt{9-14\alpha+9\alpha^2})/4$.
The perturbed part of the Hamiltonian labeled 
by $\lambda$ connects these
isolated mixed-spin singlets to form  a 2D network and thus enhances
the antiferromagnetic correlation.
We compute the staggered susceptibility $\chi_{\rm AF}$,
and  the singlet-triplet excitation gap 
$\Delta$ at the ordering wave vector.
These quantities are then  expanded as a power series in $\lambda$.
We finally  determine  the phase boundary  by 
the divergent staggered susceptibility and  the vanishing spin gap, 
which are estimated by applying  the Pad\'e approximants\cite{Pade} 
to the quantities obtained up to the finite order in $\lambda$.

To confirm  how well our mixed-spin cluster approach  works,
we first investigate the $s=1$ spin chain
without bond alternation. Performing the mixed-spin cluster expansion,
we calculate the ground state energy $E_g$, 
the staggered susceptibility $\chi_{\rm AF}$ and 
the singlet-triplet excitation gap $\Delta$ up to the eleventh, the fifth and 
the seventh order, respectively. At first sight,
the order in the series for the staggered susceptibility and the excitation gap
might not be high enough to produce the accurate values at $\lambda=1$ 
(the Haldane point) 
by means of the ordinary differential methods.\cite{Pade}  
It is remarkable, however, that  
there exists an additional symmetry property like 
$Q(\lambda)=\lambda Q(1/\lambda)$ 
for each quantity $Q$ in our effective mixed-spin chain, which 
enables us to  expand the quantity $Q$ as a power series even around 
$\lambda=1$.
Fitting this power series with that obtained by the cluster expansion,
we end up with the rather  accurate values, 
$E_g=-1.4022$, $\chi_{\rm AF}=19.6$ and $\Delta=0.404$,
which are compared with  those of the Monte Carlo simulations:
$E_g = -1.4015 \pm 0.0005$, $\Delta = 0.41$ in ref.,\cite{Night}
 and also the exact diagonalization
$\Delta = 0.411 \pm 0.001$, $\chi_{\rm AF} = 18.4 \pm 1.3$ 
in ref.\cite{Sakai1D}
\begin{figure}[htb]
\epsfxsize=7cm
\centerline{\epsfbox{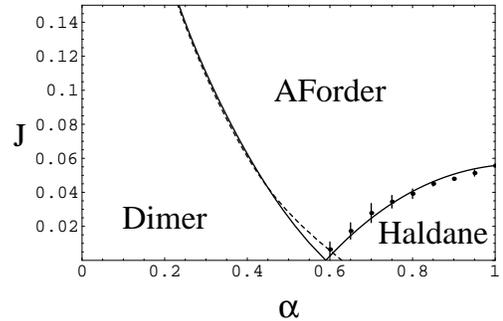}} 
\vspace{0.1cm}
\caption{Phase diagram for the 2D $s=1$ quantum spin system 
with bond alternation $\alpha$.  The phase boundary between the Haldane phase
and the ordered phase is determined by the mixed-spin cluster expansion.
The left solid (the left dashed) phase boundary around 
the dimer phase is determined by the dimer expansion with 
the biased [2/3] Pad\'e approximants 
for the excitation gap (the staggered susceptibility).
}
\label{fig:phase}
\end{figure}

To study the quantum phase transition on a 2D lattice 
by increasing the  interchain couplings,
we evaluate  the singlet-triplet excitation 
gap $\Delta$ by means of the mixed-spin 
cluster expansion up to the fifth order in $\lambda$ 
for various choices of $\alpha$ and $J$.
 It is sufficient to consider the parameter regime near  
$\lambda=1$ to discuss  the original Haldane system.
In the case without bond alternation $(\alpha=1)$,
applying the Dlog Pad\'e approximants to the spin  gap, 
the critical value $J_c =0.056\pm0.001$ and 
the critical exponent $\nu=1.86\pm0.08$ are obtained.
Our results for $\alpha=1$ are much more accurate  than
those of the mean field theory combined with the exact 
diagonalization,\cite{Sakai2D}
which claimed  the critical value  to be $J_c >0.025$.
We here note that the obtained critical exponent is different from 
the value $\nu=0.71$\cite{bias} 
expected for the 3D classical Heisenberg model.\cite{CHN}
This implies that the quantum phase transition in our generalized 
mixed-spin model does not belong to 
the universality class of the 3D classical Heisenberg model in generic cases,
although it should do in  the specific case $\lambda=1$.
Assuming that the spin gap in the vicinity of the transition point 
vanishes with the same exponent even for the Haldane system 
with bond alternation, we determine the phase boundary shown 
as the dots with error bars in Fig. \ref{fig:phase}.
The error bars come from the different values obtained 
by  different biased Pad\'e approximants employed:
[1/2], [2/1], [2/2], [2/3], [3/2] approximants. 
Since the error bars increase with the decrease of  $\alpha$ away from unity,
it seems difficult to determine the phase boundary 
in the region close to the dimer phase.
However, it is to be noted that this phase diagram should have 
the symmetry property as $J(\alpha)=\alpha J(1/\alpha)$.
Taking this into account, we can thus determine rather precisely the 
phase boundary 
between  the Haldane phase and the antiferromagnetic  phase,
which is drawn by  the solid line  in Fig. \ref{fig:phase}.
We shall see  momentarily that the critical point between the 
dimer and the Haldane phases determined in this
procedure is quite consistent  with that obtained by the dimer expansion.


Let us now turn to the dimer phase.
In this case, our mixed-spin cluster expansion is equivalent to
the standard dimer expansion.\cite{study} We perform
the dimer expansion of the staggered susceptibility 
and the spin gap up to the fifth and the sixth order 
in $\lambda$ for various $J$, respectively.
To estimate the phase boundary which separates the dimer phase and 
the antiferromagnetic  phase, we  use the ordinary Pad\'e 
approximants\cite{Pade} as well as the  biased Pad\'e approximants, 
for which  the phase transition is assumed to  belong
to the unversality class of the 3D classical Heisenberg model.\cite{CHN}
Using these Pad\'e approximants, 
we arrive at the phase diagram shown in Fig. \ref{fig:phase}.
When $J=0$ with small  $\alpha$, 
the system is reduced to the isolated $s=1$ bond-alternating chain, 
which is known to have disordered ground state with the spin gap
due to the dimer singlet. Increasing the parameter $J$ and $\alpha$, 
the antiferromagnetic correlation grows up, 
and the quantum phase transition to the magnetically ordered state occurs.
We wish to note that the critical point $(\alpha, J)=(0.59,0)$,
which is determined from the series expansion of the spin gap,
 separates the Haldane phase, the dimer phase and 
the antiferromagnetically ordered phase in Fig. \ref{fig:phase}.
Since the system in this case is reduced to 
the independent $s=1$ spin chains with bond alternation, 
our numerical results reproduce 
the well-known fact\cite{Gelone,KatoTanaka,Totsuka,Yamamoto}
that the ground state of the reduced  chain with 
$\alpha_c=0.59$ is in a critical phase
with neither the spin gap nor the long-range order.
To confirm how accurate our results for 2D cases are, we have 
directly analyzed the spin chain ($J=0$)
by applying the Dlog Pad\'e approximants to 
the spin gap computed  up to the eighth order. This gives 
$\alpha_c=0.612\pm0.004$, which is close to the value
$0.59$ obtained above, and also to $0.60 \pm 0.01$ 
obtained by DMRG.\cite{KatoTanaka}
Judging from these results, we can say that our phase boundary determined
 by the  excitation gap in Fig. \ref{fig:phase} is quite 
accurate, while that by the staggered 
susceptibility has a slight deviation only around the critical point.


\begin{figure}[htb]
\epsfxsize=7cm
\centerline{\epsfbox{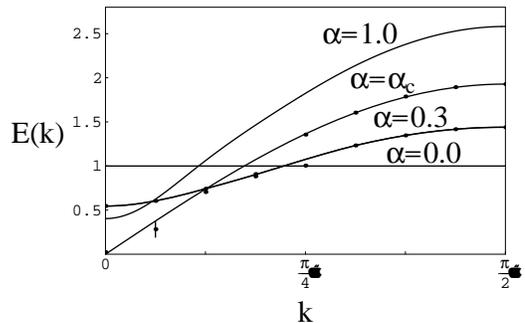}} 
\vspace{0.1cm}
\caption{Dispersion relations of the spin-triplet excited states for 
the $s=1$ chain with bond alternation $\alpha$.
The results for $\alpha=0, 0.3$ and $\alpha_c (=0.59)$ are obtained 
by the dimer expansion,
while the other for the Haldane phase is obtained by the mixed-spin 
cluster expansion with the help of a symmetry property (see the text).}
\label{fig:dispersion}
\end{figure}

In order to demonstrate that our approach is also powerful to compute
the elementary excitation with finite momenta, we show the 
calculated dispersion relation 
in Fig. \ref{fig:dispersion} along the specific line of $J=0$.
Reflecting the isolated spin-singlet structure, 
the Brillouin zone becomes half of the original one.
In the dimer phase $0<\alpha<\alpha_c$,
using the first order inhomogeneous differential method,\cite{Pade}
we can obtain the dispersion relation.
Here, to obtain the dispersion for the Haldane 
phase, we have again made use of 
the additional symmetry property inherent in the effective
mixed-spin model mentioned above.
It is to be emphasized that  such a precise 
dispersion is obtained within the lower-order perturbations, which is
indeed due to the additional symmetry we have used.

\begin{figure}[htb]
\epsfxsize=7cm
\centerline{\epsfbox{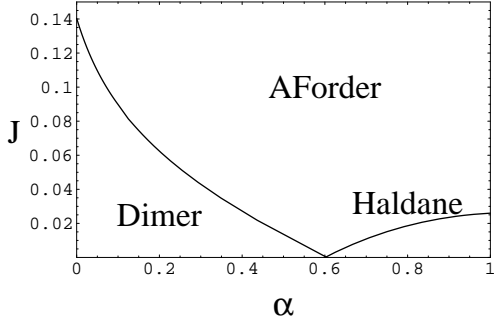}} 
\vspace{0.1cm}
\caption{Phase diagram for the 3D  $s=1$ spin system
with bond alternation $\alpha$.}
\label{fig:phase3D}
\end{figure}

We now move to the 3D system.
The advantage of our approach is particularly remarkable
for the 3D problem because
other numerical methods may often meet some difficulties
to treat a large system in the 3D case. 
We here consider a cubic lattice system 
by adding the interchain couplings $J$ in the $z$ direction 
to the spin model discussed above.
By extending our treatment to the 3D system,
we thus study the competition between the two kind of gapped states and 
the antiferromagnetic state.  Applying the dimer expansion to
calculate the spin gap and the staggered susceptibility 
up to the fifth order and using the Dlog [2/2] Pad\'e approximants,
we first  determine the phase boundary which separates the dimer and 
the antiferromagnetic ordered phase in Fig. \ref{fig:phase3D}.
When $\alpha=0$, our system reproduces the $s=1$ bilayer Heisenberg model.
Increasing the inter-dimer coupling $J$ from zero, 
the antiferromagnetic correlation grows up and 
the quantum phase transition occurs at $J_c=0.143\pm0.006$.
We note that the quantum phase transitions
in the bilayer model have already been studied 
by Gelfand et al\cite{bilayerS}  
with the series expansion method.
On the other hand, 
to observe the phase transition from  the Haldane phase
to the ordered phase, we further  perform the mixed-spin 
cluster expansion up to the fourth order for both of the 
above quantities.  In the homogeneous case $(\alpha=1)$, 
by analyzing the data in terms of various Dlog Pad\'e approximants 
we end up with  the critical point $J_c=0.026\pm 0.001$,
which is consistent with those of 
the non-linear $\sigma$ model approach\cite{Senechal} and
the mean field theory combined with the numerical method.\cite{Sakai2D}
The phase diagram thus determined is shown in Fig. \ref{fig:phase3D}.

In summary, we have investigated the quantum phase transitions
for the $s=1$ quantum systems with the 2D and 3D structures.
Using the series expansion,
we have discussed how the dimer phase and the Haldane phase 
realized in 1D  compete with 
the magnetically ordered phase in higher dimensions.
 In particular, we have proposed 
a novel approach based on the mixed-spin cluster expansion which realizes  
the idea of the VBS in the perturbation theory.
This new approach has made it possible to treat 
the Haldane phase in the series expansion framework,
which  was not dealt with so far by ordinary series 
expansion methods. For the spin chain case, 
we have obtained fairly good results comparable to
other numerical methods.\cite{Night,Sakai1D}  For the 2D and 3D
cases, the phase diagram has been
determined rather precisely by making 
use of an additional symmetry property in the effective mixed-spin model.
It is quite interesting to futher apply the mixed-spin cluster approach
to the frustrated case, the anisotropic case, etc., 
in quasi-1D Haldane systems,
which is now under consideration.

The work is partly supported by a 
Grant-in-Aid from the Ministry of Education, Science, Sports, and Culture.
A. K. is supported by the Japan Society for the Promotion of Science.
A part of numerical computations in this work was carried out 
at the Yukawa Institute Computer Facility.




\begin{thebibliography}{99}
\bibitem{comment}
If we further divide all $s=1$ spins,  we end up with
the 2-leg ladder with the diagonal bonds 
without the rung bonds. This was previously pointed out by 
M. P. Gelfand:
Phys. Rev. B {\bf 43} 8644 (1991) and also by 
S. R. White:
Phys. Rev. B {\bf 53} 52 (1996).
\end{thebibliography}
\end{document}